\documentclass[a4paper,pre,nofootinbib,twocolumn,superscriptaddress]{revtex4-2}
\usepackage{amssymb,amsmath,amsthm,physics,bm,extarrows,appendix,color,graphicx,hyperref}
\allowdisplaybreaks[4]

\begin{document}

\title{
Observable-manifested correlations in many-body quantum chaotic systems
}

\author{Xiao Wang}\email{wx2398@mail.ustc.edu.cn}
\affiliation{ Department of Modern Physics, University of Science and Technology of China,
Hefei 230026, China}
\affiliation{CAS Key Laboratory of Microscale Magnetic Resonance, Hefei 230026, China}

\author{Jiaozi Wang}
\affiliation{Department of Mathematics/Computer Science/Physics, University of Osnabr\"uck, D-49076
Osnabr\"uck, Germany}

\author{Wen-ge Wang}\email{wgwang@ustc.edu.cn}
\affiliation{ Department of Modern Physics, University of Science and Technology of China,
	Hefei 230026, China}
\affiliation{CAS Key Laboratory of Microscale Magnetic Resonance, Hefei 230026, China}
\affiliation{Anhui Center for fundamental sciences in theoretical physics, Hefei 230026, China}

\date{\today}

\begin{abstract}
    In this paper, we investigate the distinctions between realistic quantum chaotic
    systems and random models from the perspective of observable properties, particularly
    focusing on the eigenstate thermalization hypothesis (ETH). Through numerical
    simulations, we find that for realistic systems, the envelope function of off-diagonal elements of observables exhibits an
    exponential decay at large $\Delta E$, while for randomized models, it tends to be flat.
We demonstrate that the correlations of chaotic eigenstates, originating from the delicate structures of Hamiltonians, play a crucial role in the non-trivial structure of the envelope function.
Furthermore, we analyze
the numerical results from the perspective of the dynamical group elements in Hamiltonians.
    Our findings highlight the
    importance of correlations in physical chaotic systems and provide insights into the
    deviations from RMT predictions. These understandings offer valuable directions
    for future research.
\end{abstract}

\maketitle

\section{Introduction}\label{Sec_Introduction}

What is the most intuitive perception of chaotic motion?
A widely accepted understanding
 involves envisioning chaotic motion as a type of motion resulting from intricate
interactions characterized by significant disorder and randomness.
Indeed, disorder and randomness are significant characteristics of quantum chaotic systems \cite{Haake-book}.
In particular, similarity of statistical properties of quantum systems to
predictions of the random matrix theory
(RMT) has long been used as an indicator of quantum chaos
\cite{Berry-chaos-81,Casati-chaos80,Berry-chaos-85,Sieber-chaos-01,Haake-book,Kaufman-chaos-79,Mueller-chaos-04,Mueller-chaos-05,Wigner-chaos}.
Moreover, it has also been found that eigenstates of quantum chaotic systems exhibit universal
properties, with their rescaled components on certain bases following the Gaussian distribution
\cite{Berry_1977,pre18-EF-BC,Buch-EF-82,Benet-EF-03,Benet-EF-07,Meredith-EF-88,PhysRevLett.71.1291-Rigol-Entropy},
in consistency with RMT.

However, disorder and randomness do not fully capture the essence of quantum chaos.
Despite the similarity in fluctuation properties described above, quantum chaotic systems
deviate from fully random systems described by RMT in various ways.
For example, it is well known that
average properties, such as averaged spectral density and averaged shape of eigenfunctions
(on a given basis), are usually system-dependent and do not show any universal
behavior, deviating from RMT.

In this paper, we study distinctions between quantum chaotic systems and
RMT from the viewpoint of observable properties,
particularly that stated in the framework of the eigenstate thermalization hypothesis (ETH)
\cite{Rigol-AiP16, Deutch91, srednicki1994chaos, srednicki-JPA96,
srednicki1999approach, RS-PRL12, Srednicki_Rigol_2013, DePalma_PRL_2015, Deutch-RPP18,
Turner_2018}.
 For observables $O$ on the eigenbasis of the system's Hamiltonian $H$, the ETH
ansatz conjectures that
\begin{equation}\label{ETH}
O_{ij}=\bra{E_i}O\ket{E_j}=O(E_i)\delta_{ij}+f(E_i,E_j)r_{ij},
\end{equation}
where $E_j$ and $\ket{j}$ denote eigenvalues and eigenstates of $H$, respectively.
Here, $O(E)$ and $f(E_i,E_j)$ are smooth functions of their arguments, $\delta_{ij}$ is the
Kronecker Delta function, and $r_{ij}=r^*_{ji}$ are random variables with a normal
distribution (zero mean and unit variance). Although the ETH remains a hypothesis due to
the lack of rigorous proof, most aspects of the ETH have been confirmed by numerical
simulations \cite{Rigol-AiP16, Deutch-RPP18, BMH-PRE14, BMH-PRE15, DLL-PRE18, Vidmar_2019,
Vidmar21-PRB, YWW2022PS}. It is now widely accepted that the ETH holds, at least, for quantum
chaotic systems when considering few-body observables.

 As is known, for models associated with realistic objects,
 the envelope function $f(E_i, E_j)$ in ETH significantly deviates
from RMT predictions \cite{Rigol-AiP16,PRA86-Feing-Peres,Wilkinson_1987,Prosen_1994,Muller_1995,Srednicki_PRE1998,YWW2022PS,WangXiao_2024}.
 We are to show that this deviation stems from
correlations in chaotic energy eigenstates, which in turn originate from delicate
structures of the Hamiltonians.
In particular, such structures are to be studied in the perspective of
the underlying dynamical group of the Hamiltonian.
 In addition, a system-environment uncoupled basis is to be employed for
 giving certain explanations to the deviations found.

The rest of the paper is organized as follows. In Sec.\ref{Sec_Model}, examples are given,
illustrating the variation of the envelope function $f(E_i, E_j)$ when the Hamiltonians
are changed. In Sec.\ref{Sec_Explanations}, we analyse the numerical result presented in
Sec.\ref{Sec_Model} from three different aspects. Finally, conclusions and some
discussions are given in Sec.\ref{Sec_Conclusion}.

\section{Numerical simulations for the envelope function $f(E_i,E_j)$}\label{Sec_Model}

\subsection{In a defect Ising chain}

We begin with presenting examples of the envelope function $f(E_i,E_j)$.
In the following, the eigenstates and eigenvalues of Hamiltonian $H$ are denoted by
$\ket{E_j}$ and $E_j$, respectively:
\begin{equation}
    H\ket{E_j}=E_j\ket{E_j}.
\end{equation}
According to the definition in Eq.(\ref{ETH}), the envelope function $f(E_i,E_j)$ is
obtained by taking average of the off-diagonal elements,
\begin{equation}\label{Expression_g2}
    f^2(E_i, E_j)=\overline{\abs{O_{ij}^{\rm off}}^2}:=\overline{\abs{\bra{E_i}O\ket{E_j}}^2}
\end{equation}
 for $E_i\neq E_j$, over narrow energy shells around $E_i$ and
$E_j$. In our numerical calculations, each energy shell contains
approximately 15 levels around $E_i$ and $E_j$.

 As the first model, we consider the defect Ising chain (DIS), which consists of $N$
$\frac{1}{2}$-spins subjected to an inhomogeneous transverse field. The Hamiltonian
is given by:
\begin{equation}\label{DIS_Hamiltonian}
    \begin{aligned}
        H_{\rm DIS}=\frac{B_x}{2}\sum_{l=1}^N &\sigma^l_x + \frac{d_1}{2} \sigma^1_z + \frac{d_5}{2} \sigma^5_z\\
        &+\frac{J_z}{2}\left(\sum_{l=1}^{N-1}\sigma^l_z \sigma^{l+1}_z+\sigma^N_z \sigma^1_z\right),
    \end{aligned}
\end{equation}
where $\sigma^l_{x,y,z}$ are Pauli matrices at site $l$.
 The parameters are set as
$B_x=0.9$, $d_1=1.11$, $d_5=0.6$, and $J_z=1.0$. The number of spins in the system is
$N=14$. Under these parameters, the system is chaotic.

In this paper, the spin direct product basis of $N$ $\frac{1}{2}$-spins will be
denoted by $\ket{\alpha}$, which represents the common eigenstate of all $\{\sigma^l_z\}$.
For instance, one such $\ket{\alpha}$ can be expressed as:
\begin{equation}
    \ket{\alpha}=\ket{\uparrow}_1\otimes\ket{\downarrow}_2\otimes\cdots\otimes\ket{\uparrow}_N,
\end{equation}
where $\ket{\uparrow}_l$ and $\ket{\downarrow}_l$ are eigenstates of $\sigma^l_z$.

\begin{figure}[t!]
    \centering
    \includegraphics[width=1\linewidth]{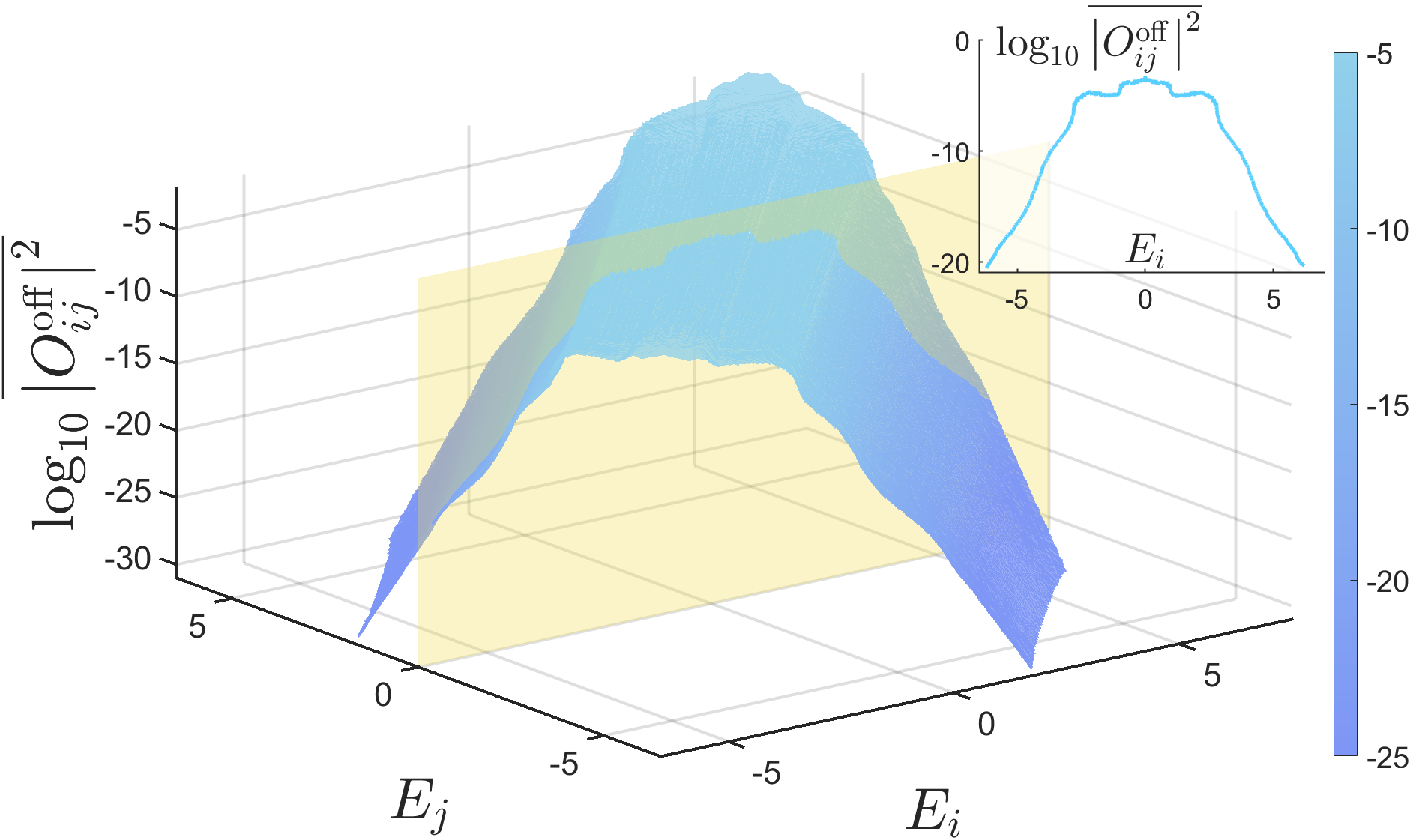}
    \caption{$\log_{10}\overline{\abs{O_{ij}^{\rm off}}^2}$ versus $(E_i, E_j)$ in the DIS model for the observable $O=\sigma_x^7$. The inset shows a cross-section
    taken at $E_j=-0.0013$ (indicated by the yellow plane).
     }
    \label{g2_DIS-Sx7}
\end{figure}

\begin{figure}[t!]
    \centering
    \includegraphics[width=1\linewidth]{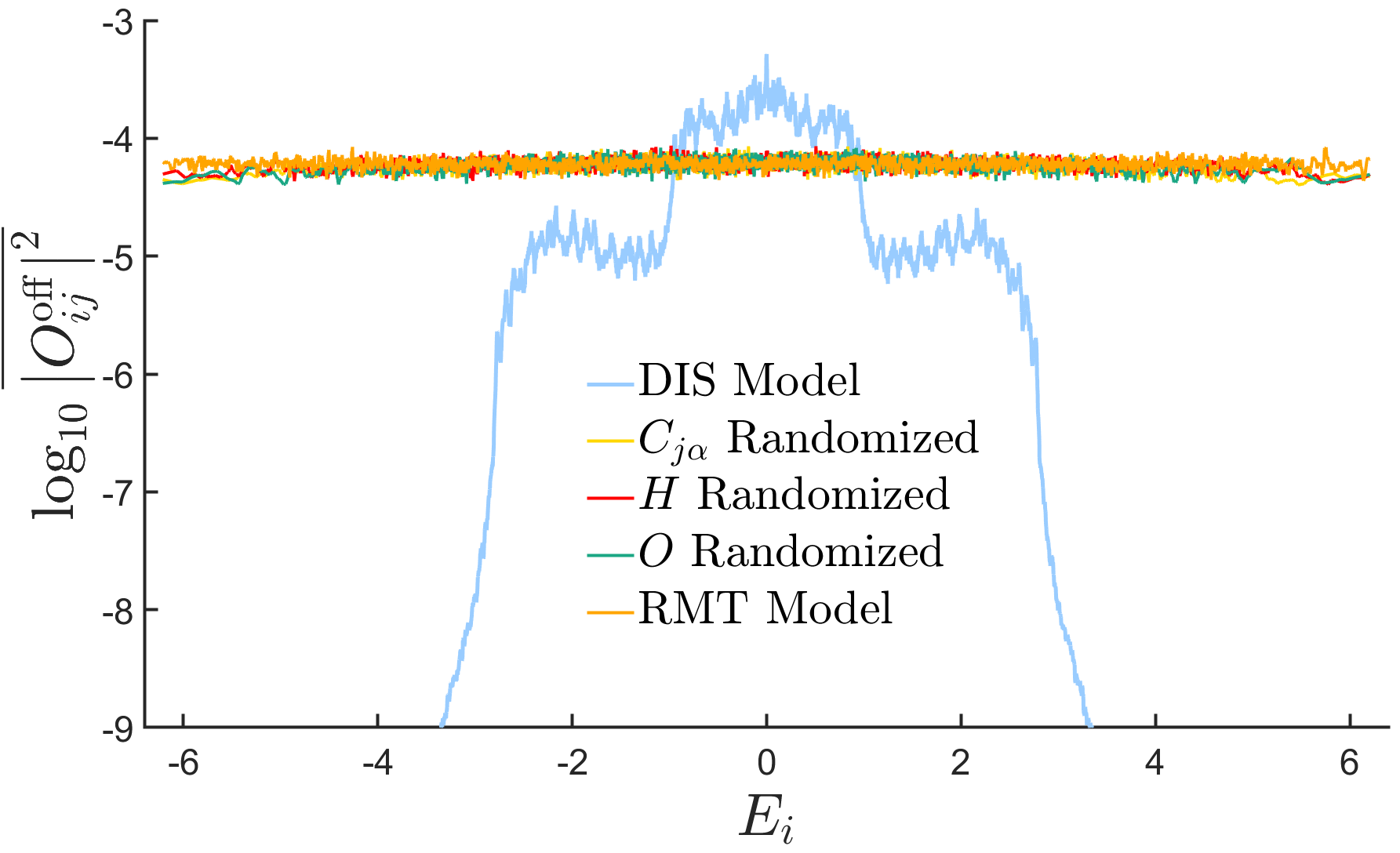}
    \caption{$\log_{10}\overline{\abs{O_{ij}^{\rm off}}^2}$ in different cases.
    The blue line is an enlargement of the cross-section shown in Fig.\ref{g2_DIS-Sx7}.
    The yellow line represents a cross-section of $\log_{10}\overline{\abs{\bra{E_i^{(R)}}O\ket{E_j^{(R)}}}^2}$,
    where $\ket{E_j^{(R)}}$ is defined in Eq.(\ref{eq-DISRW}).
    The red line depicts a cross-section of $\log_{10}\overline{\abs{O_{ij}^{\rm off}}^2}$
    using the randomized Hamiltonian $H^{(R)}$ defined in Eq.(\ref{Randomized_H}).
    In all of above cases, the observable $O$ is taken as $O=\sigma^x_7$.
    The green line shows a cross-section of $\log_{10}\overline{\abs{\bra{E_i}O^{(R)}\ket{E_j}}^2}$,
    with $O^{(R)}$ defined in Eq.(\ref{Randomized_O}).
    Lastly, the orange line corresponds to the prediction of GOE.
    All these cross-sections are taken with $E_j$ fixed at the centers of the spectra.}
    \label{DIS_Compare}
\end{figure}

Fig.\ref{g2_DIS-Sx7} shows $f^2(E_i, E_j)$ as a function of $(E_i, E_j)$ for the
observable $O=\sigma_x^7$. For a clear sight, we show a cross-section at a fixed value of $E_j$ in
the inset of Fig.\ref{g2_DIS-Sx7},
where a slowly changing
plateau is seen at small energy differences $\Delta E:=\abs{E_i-E_j}$, followed by an exponential
decay at large $\Delta E$.
 In fact, as is known in numerical simulations,
 a slowly changing plateau at small $\Delta E$, which is followed by
 an exponential decay at large $\Delta E$,
 is a typical behavior of the $f^2(E_i, E_j)$ function in quantum chaotic systems
 \cite{Rigol-AiP16,Rigol_2008,
 Srednicki_Rigol_2013,Rigol_2021,Rigol_2020,Rigol_2017,Vidmar_2019,Rigol_2019,
 Gemmer_2020,YWW2022PS,WangXiao_2024}.
 In contrast, in a fully random model whose Hamiltonian
 matrix is a typical element of the Gaussian Orthogonal Ensemble (GOE),
 the envelope function $f^2(E_i,E_j)$ is flat
 (as shown in Fig.\ref{DIS_Compare}), without any exponential decay \cite{Rigol-AiP16}.

\begin{figure}[t!]
    \centering
    \includegraphics[width=1\linewidth]{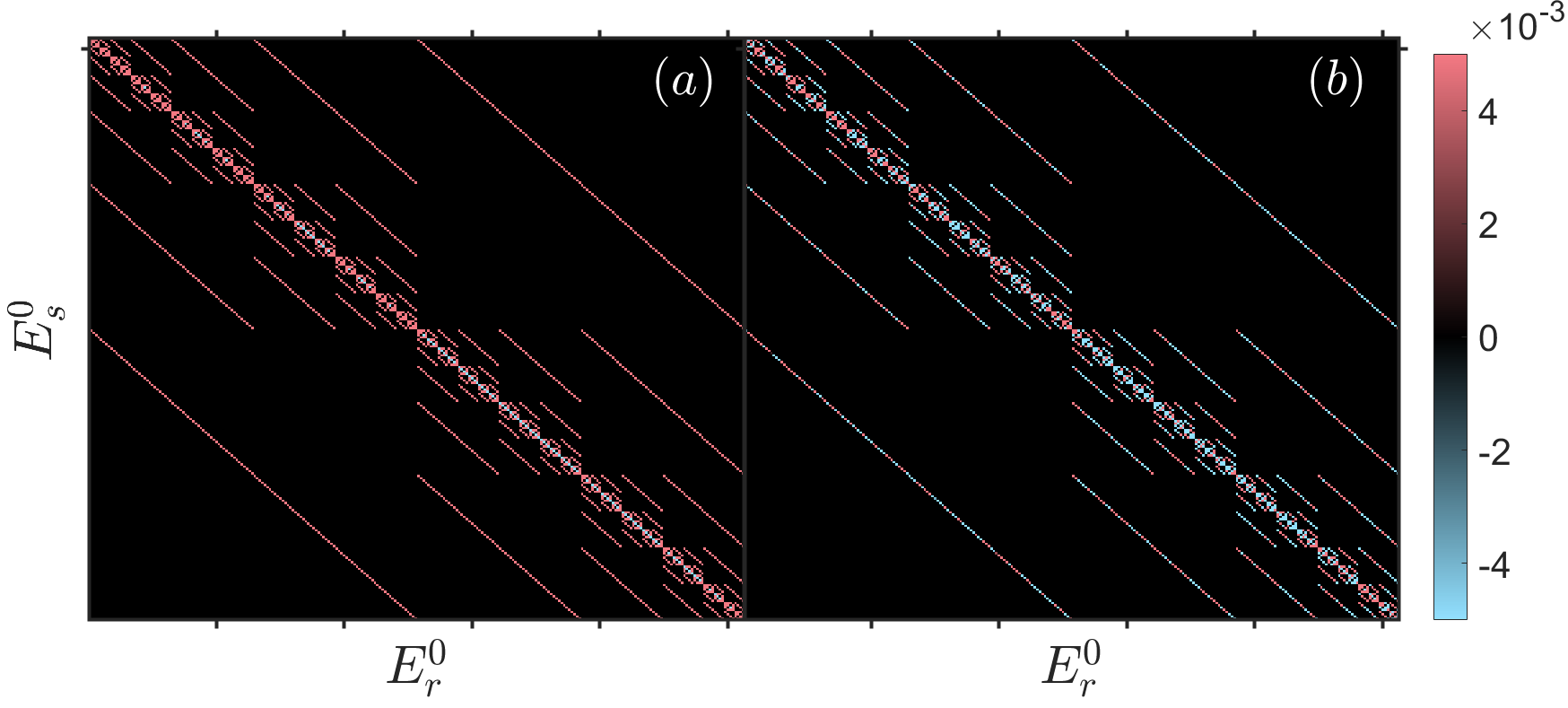}
    \caption{The matrix elements of the original DIS Hamiltonian $H_{\rm DIS}$ (defined in
    Eq.(\ref{DIS_Hamiltonian})) and the randomized DIS Hamiltonian $H^{(R)}$ (defined in
    Eq.(\ref{Randomized_H})) in the spin direct product basis $\ket{\alpha}$. Specifically,
    panel (a) illustrates the matrix elements $\bra{\alpha}H_{\rm DIS}\ket{\beta}$,
    while panel (b) shows the matrix elements $\bra{\alpha}H^{(R)}\ket{\beta}$. Here,
    the total spin number $N=8$.
    }
    \label{DIS_HShape_Alpha}
\end{figure}

It is worth noting that not just GOE random matrix Hamiltonian can produce eigenstates
with disruption of correlations, sufficient to flatten the $f^2(E_i, E_j)$ function.
The red line in Fig.\ref{DIS_Compare} shows the $f^2(E_i, E_j)$ of another system, where
the observable $O$ is again taken as $O=\sigma_x^7$, but the Hamiltonian $H$ is randomized
from the original Hamiltonian of the DIS model as follows.
That is, the randomized Hamiltonian
$H^{(R)}$ is generated by the following method:
\begin{equation}\label{Randomized_H}
    \bra{\alpha}H^{(R)}\ket{\beta}=r_{\alpha\beta}\bra{\alpha}H_{\rm DIS}\ket{\beta},
\end{equation}
where $r_{\alpha\beta}=r_{\beta\alpha}$ are independent random numbers drawn from a
Gaussian distribution. This operation retains all zero elements of the matrix $\bra{\alpha}H\ket{\beta}$,
as well as the average magnitudes of $\abs{\bra{\alpha}H\ket{\beta}}$.
 In other words, the main structural features of $H_{\rm DIS}$
are preserved (as shown in Fig.\ref{DIS_HShape_Alpha}).

Note that, since the original Hamiltonian $H_{\rm DIS}$ of the DIS model has a sparse
matrix in the $\ket{\alpha}$-representation, the number of random parameters contained
in $H^{(R)}$ is much less than that in a GOE random matrix. However, despite retaining
the main structural features of $H_{\rm DIS}$ and containing far less random parameters than a
GOE random matrix, Fig.\ref{DIS_Compare} shows that these random parameters are still
enough to disrupt the correlations between eigenstates and observables and further flatten the $f^2(E_i, E_j)$
function.

\subsection{For two types of Hamiltonians}\label{Sec_IncreaseRandomness}

In this section, we show that the behavior of the envelope function $f(E_i,E_j)$
is closely related to ``realisticity" of the Hamiltonian. Concretely, it is shown that,
when the Hamiltonian contains only local interactions involving adjacent
particles, the $f(E_i,E_j)$ functions  have an exponential decay at large
$\Delta E$. On the contrary, once numerous non-local interactions enter the Hamiltonian,
the exponential decay of $f(E_i,E_j)$ will disappear.
For this purpose, we  are to study two types of Hamiltonians.

In the first type of Hamiltonian, indicated as $H^{\rm DIS}_{n}$,
is obtained by adding second-neighboring interaction and so on to the DIS Hamiltonian.
More exactly, it is written as
\begin{equation}
    H^{\rm DIS}_{n} = H_{\rm DIS} + \sum_{k=1}^n V_k,
\end{equation}
where $V_1=0$, and $V_k\ (k\geqslant 2)$ represents the sum of all adjacent $n$-point interactions along $x$ direction.
For example,
\begin{subequations}
    \begin{align}
        V_2&=\sum_{l=1}^N J^{l,(l+1)}_x \sigma^l_x \sigma^{l+1}_x,\\
        V_3&=\sum_{l=1}^N J^{l,(l+1),(l+2)}_x \sigma^l_x \sigma^{l+1}_x \sigma^{l+2}_x,\\
        &\cdots\cdots \notag
    \end{align}
\end{subequations}
In the above expressions,  modulo $N$ is taken for indices exceeding $N$, and the coefficients
$\{J^{l,(l+1)}_x,\ J^{l,(l+1),(l+2)}_x\}$ are independent Gaussian random numbers with mean zero.
Thus, by definition, $H^{\rm DIS}_{n}$ only contain local interactions.

\begin{figure}[t!]
    \centering
    \includegraphics[width=1\linewidth]{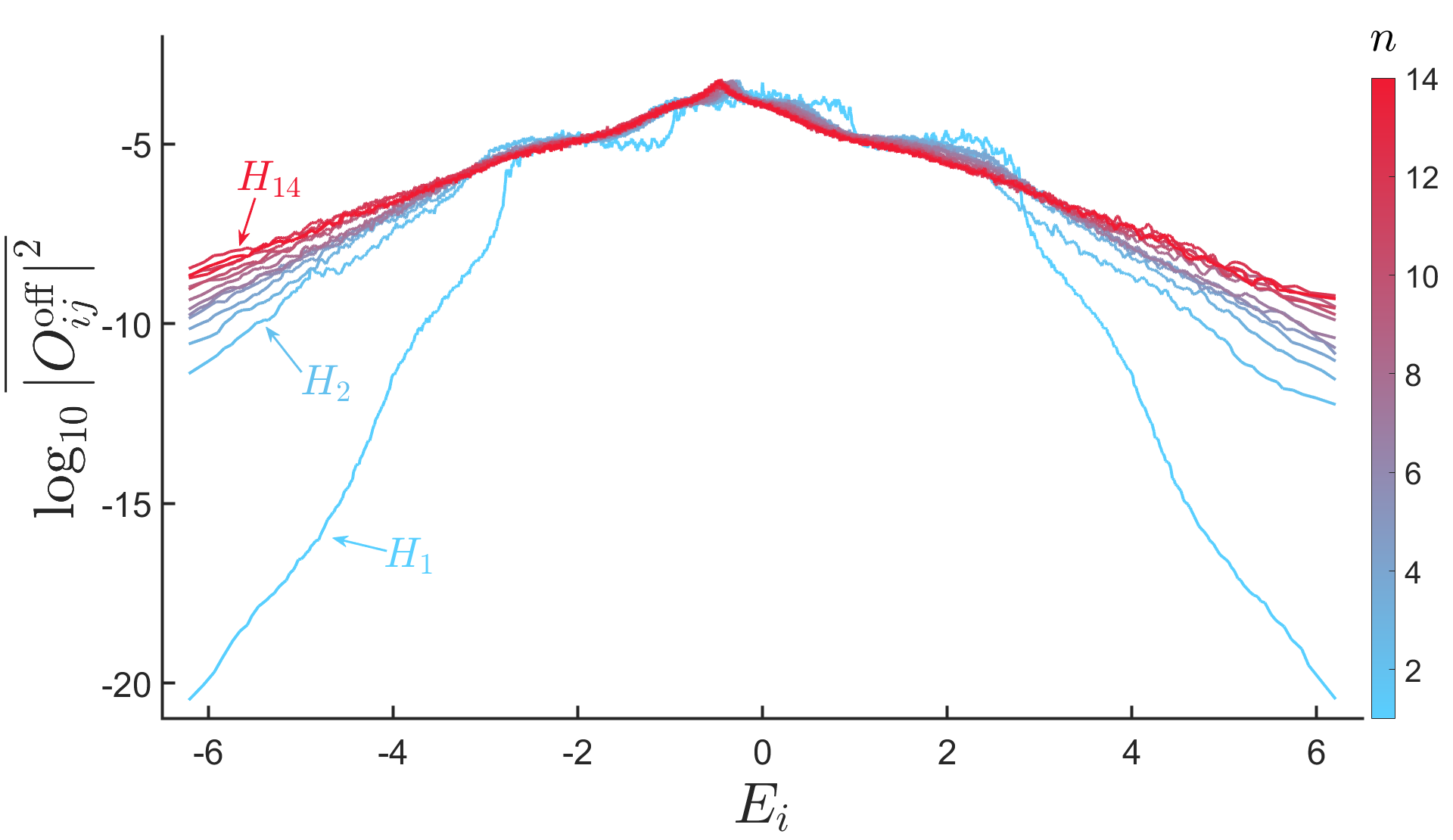}
    \caption{$\log_{10}\overline{\abs{O_{ij}^{\rm off}}^2}$ computed using modified DIS
    models, which incorporate varying numbers of independent parameters in their
    Hamiltonians $H_n$. $N_R$ represents the number
     of independent parameters. Under all
    these conditions, the observable $O$ is consistently set as $O=\sigma_x^7$. For all
    cross-sections, the energy $E_j$ is fixed at the central value of the respective
    spectra.}
    \label{g2_DIS_ActionLength}
\end{figure}

Fig.\ref{g2_DIS_ActionLength} depicts behaviors of $f^2(E_i, E_j)$ obtained from
different Hamiltonians $H^{\rm DIS}_{n}$. The observable $O$ is also set as $O=\sigma_x^7$.
It can be seen that an exponential decay always exists.

It's deserved to point out that, although the interacting strength coefficients
$\{J^{l,(l+1)}_x,\ J^{l,(l+1),(l+2)}_x,\cdots\}$ are taken as random numbers in our
numerical calculations, the behavior of $f^2(E_i, E_j)$ are actually not sensitive to
the randomness of these coefficients. Even if $\{J^{l,(l+1)}_x,\ J^{l,(l+1),(l+2)}_x,\cdots\}$
are taken as equal constants, the result will be similar to that shown in Fig.\ref{g2_DIS_ActionLength}.

 The purpose of studying a second type of system, whose Hamiltonian is indicated as $H^{(R)}_{N_R}$,
 is to give further study for effects of the randomization introduced to
 the Hamiltonian $H^{(R)}$ in Eq.(\ref{Randomized_H}).
 For this purpose, we divide the set of the elements of $H_{\rm DIS}$ into
 $N_R$ subsets, which possess equal number of elements,
 and, then, multiply each subset by a random number.
 For example, in the case of  $N_R=2$, a first half of the matrix of $H_{\rm DIS}$ is multiplied by a random number,
 meanwhile, the second half is multiplied by another random number.
 Note that the above procedure does not change zero elements of the matrix of $H_{\rm DIS}$.
 And, when $N_R$ reaches
its maximum value $N_R=2^{N-1}\times N$, $H^{(R)}_{N_R=2^{N-1}\times N}$ will be the same
as the randomized Hamiltonian $H^{(R)}$ in Eq.(\ref{Randomized_H}).

\begin{figure}[t!]
    \centering
    \includegraphics[width=1\linewidth]{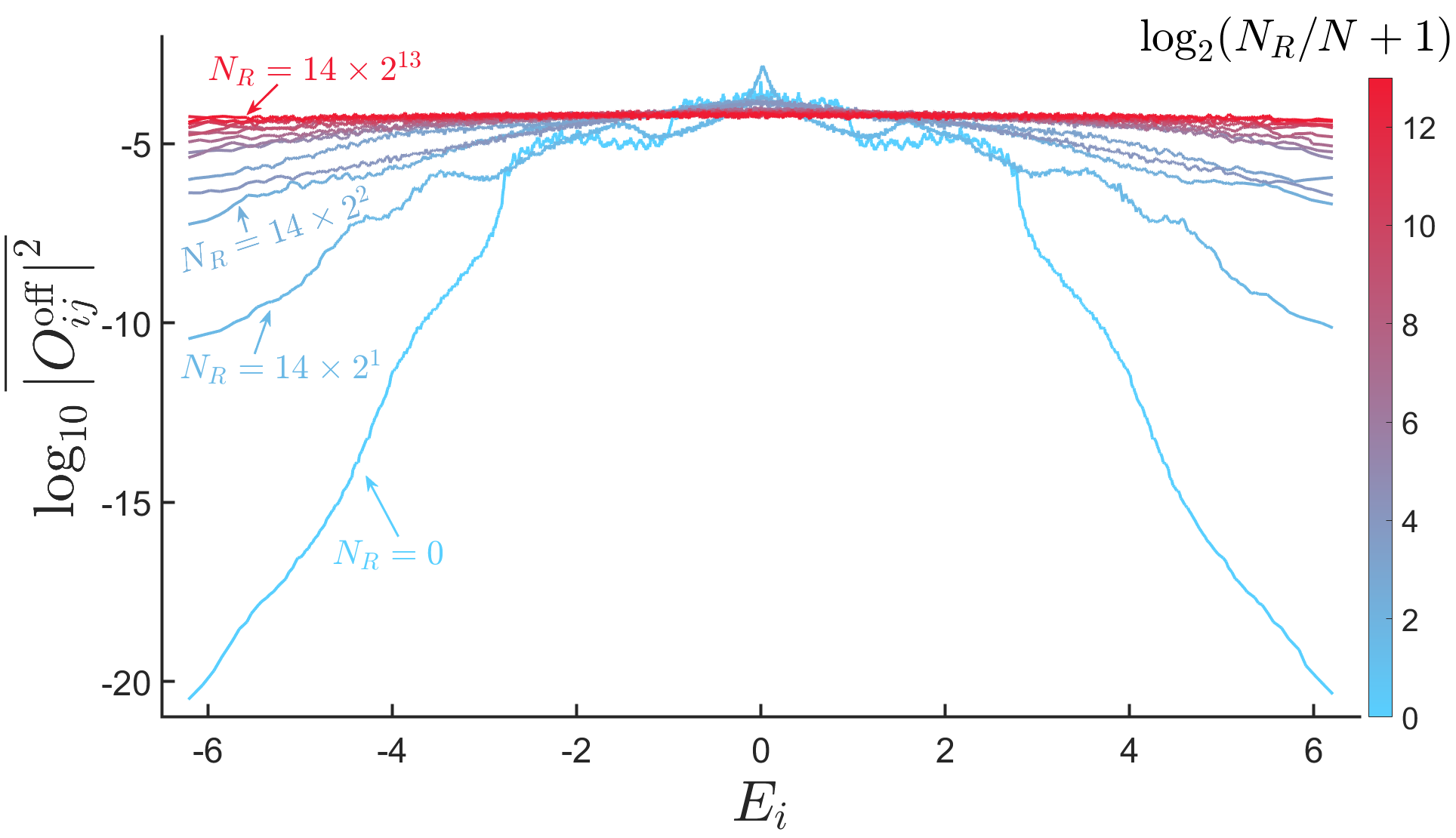}
    \caption{$\log_{10}\overline{\abs{O_{ij}^{\rm off}}^2}$ computed using modified DIS
    models, which incorporate varying numbers of independent parameters in their
    Hamiltonians by the method shown in the Appendix.
    $N_R$ represents the number of independent parameters. Under all these conditions,
    the observable $O$ is consistently set as $O=\sigma_x^7$. For all cross-sections,
    the energy $E_j$ is fixed at the central value of the respective spectra.}
    \label{g2_DISC}
\end{figure}

Fig.\ref{g2_DISC} depicts the behavior of $f^2(E_i, E_j)$
obtained from different $H^{(R)}_{N_R}$.
It shows that, with increase of $N_R$,
the exponential-decay behavior of $f^2(E_i, E_j)$ is gradually disrupted.
This is in contrast to what has been observed in the first type of Hamiltonian.

\section{Further understanding for the numerical results}\label{Sec_Explanations}

In this section, we give discussions, which are useful for understanding
numerical results discussed above.

\subsection{Correlations in Chaotic Eigenfunctions}\label{Sec_Correlation}

In this section, we study connection between the exponential decay of $f^2(E_i, E_j)$
at large $\Delta E$, which has been discussed above,
and correlations between eigenstates and the observable.

To this end, We expand the energy eigenstates $\ket{E_j}$ of the DIS model in the spin direct product basis $\{\ket{\alpha}\}$
 (which is also the eigenbasis of the observable of interest $\sigma^l_z$)
as follows:
\begin{equation}\label{StateEj_Expansion}
    \ket{E_j}=\sum_\alpha C_{j\alpha}\ket{\alpha},
\end{equation}
 where $C_{j\alpha}$ are real numbers.
 We are also to study ``randomized'' wavefunctions," denoted by $\ket{E_j^{(R)}}$,
\begin{equation}\label{eq-DISRW}
    \ket{E_j^{(R)}}    =\sum_{\alpha}e^{i\theta_{j\alpha}^{(R)}}\abs{C_{j\alpha}}\ket{\alpha},
\end{equation}
 where $\theta_{j\alpha}^{(R)}$ are randomly chosen from the two values of $0$ and $\pi$.
This operation preserves the magnitude of the eigenfunction $C_{j\alpha}$ while disrupting
the phase correlations among the components of the wavefunctions manually.

Based on the above construction, we have calculated the matrix elements $\overline{\abs{\bra{E_i^{(R)}}O\ket{E_j^{(R)}}}^2}$,
where $O$ is again taken as $O=\sigma_x^7$. A cross-section of the result is also plotted
in Fig.\ref{DIS_Compare} (yellow line).

Fig.\ref{DIS_Compare} shows that, in $\overline{\abs{\bra{E_i^{(R)}}O\ket{E_j^{(R)}}}^2}$,
the exponential decay at large $\Delta E$ disappears, and the $f^2(E_i, E_j)$ function
becomes similar to the result predicted by the random matrix model. This finding indicates
that the correlations among the phases of the original eigenfunctions of the DIS model
are crucial for maintaining the exponential decay of the $f^2(E_i, E_j)$ function. When
these correlations are destroyed, the $f^2(E_i, E_j)$ function becomes structureless.

Besides the conditions discussed above, we would also like to point out that randomization
of the observable $O$ can also flatten the $f^2(E_i, E_j)$ function. The green line in
Fig.\ref{DIS_Compare} shows the shape of $\overline{\abs{\bra{E_i}O^{(R)}\ket{E_j}}^2}$,
where $\ket{E_{i/j}}$ are energy eigenstates of the original DIS model, and the randomized
observable $O^{(R)}$ is constructed as follows:
\begin{equation}\label{Randomized_O}
    \bra{\alpha}O^{(R)}\ket{\beta}=r_{\alpha\beta}\bra{\alpha}\sigma_x^7\ket{\beta}.
\end{equation}
The $r_{\alpha\beta}=r_{\beta\alpha}$ are also independent random numbers drawn from a
Gaussian distribution. From Fig.\ref{DIS_Compare}, we can see that in this case, the
behavior of the $f^2(E_i, E_j)$ function is again close to that in the random matrix
model but far from the rapid decay behavior in the original DIS model.

The above numerical simulations show that correlations in energy
eigenstates and observables are crucial for the non-trivial structure of the envelope
function $f(E_i,E_j)$.
In particular, $f(E_i,E_j)$ becomes flat (namely, structureless), once such correlations
are destroyed.

\subsection{Relevance of Dynamical Group}

The numerical results presented in the preceding sections indicate that strong
correlations in chaotic eigenfunctions are closely related to the ``realisticity" of the
Hamiltonian. In this section, we discuss in the perspective of the
so-called dynamical group.

As is known, in a model related to realistic objects, the Hamiltonian $H$ is certain
function of the generators of some Lie group, known as the dynamical group.
For instance, consider a system involving $N$ $\frac{1}{2}$-spin particles. The
Hamiltonian for such a system is a function of operators structured as follows:
\begin{equation}\label{SU2N_Group}
    g=g^1\otimes g^2 \otimes\cdots\otimes g^l\otimes\cdots\otimes g^N,
\end{equation}
where $g^l$ represents one of the four possible operators:
\begin{equation}\label{SU2_Group}
    g^l=\sigma^l_x,\sigma^l_y,\sigma^l_z,\text{or }I^l,
\end{equation}
with $\sigma^l_{x,y,z}$ the Pauli matrices and $I^l$ the identity
operator at site $l$. The four operators in Eq.(\ref{SU2_Group}) are the four generators
of the $SU(2)$ group, while the $g$ operators in Eq.(\ref{SU2N_Group}) are generators
of the group $[SU(2)]^N$,
\begin{equation}
    [SU(2)]^N:=\underbrace{SU(2)\otimes SU(2)\otimes\cdots\otimes SU(2)}_{\text{Direct product of $N$ groups}}.
\end{equation}

From a physical viewpoint, each group generator $g$ signifies a particular kind of
interaction among particles within the system. For example,
\begin{equation}
    g=\sigma^l_x\otimes \sigma^l_x \otimes I^3 \otimes\cdots\otimes I^N
\end{equation}
represents the interaction between the first and second spins.
Independent parameters mentioned above refers to coefficients of those generators
that are used in the construction of the Hamiltonian.

Among all the above discussed $g$-operators, the majority represent non-local interactions.
In other words, $g$-operators that can be included in realistic models only account for a small portion,
whose number is far less than the dimension of the Hilbert space.
In such models, all the system's
properties, including its eigenstates, should in fact depend only on a small number of
generators (operators). Generically, this may imply strong correlations within the
eigenstates.

For example, the DIS Hamiltonian $H_{\rm DIS}$ in Eq.(\ref{DIS_Hamiltonian})
is described by $(2N+2)$ dynamical group generators,
meanwhile, as discussed previously, the DIS energy eigenstates exhibit strong correlations,
and the corresponding $f(E_i,E_j)$ function have exponential behavior.
In contrast, the Hamiltonian of the RMT model incorporates all $4^N$ combinations of
the generator $g$, most of which correspond to quite complex interactions,
which disrupted correlations within the eigenstates, and the $f(E_i,E_j)$ function becomes
flatten.
\color{black}

 Moreover, usually, physical observables $O$ are also generated from
 a small number of generators of the dynamical group.
 Indeed, numerical simulations discussed previously show strong correlations between
 such physical observables $O$ and the DIS Hamiltonian.

\subsection{Explanations in uncoupled Representation}\label{Sec_UncoupledBasis}

\begin{figure}[h!]
    \centering
    \includegraphics[width=0.962\linewidth]{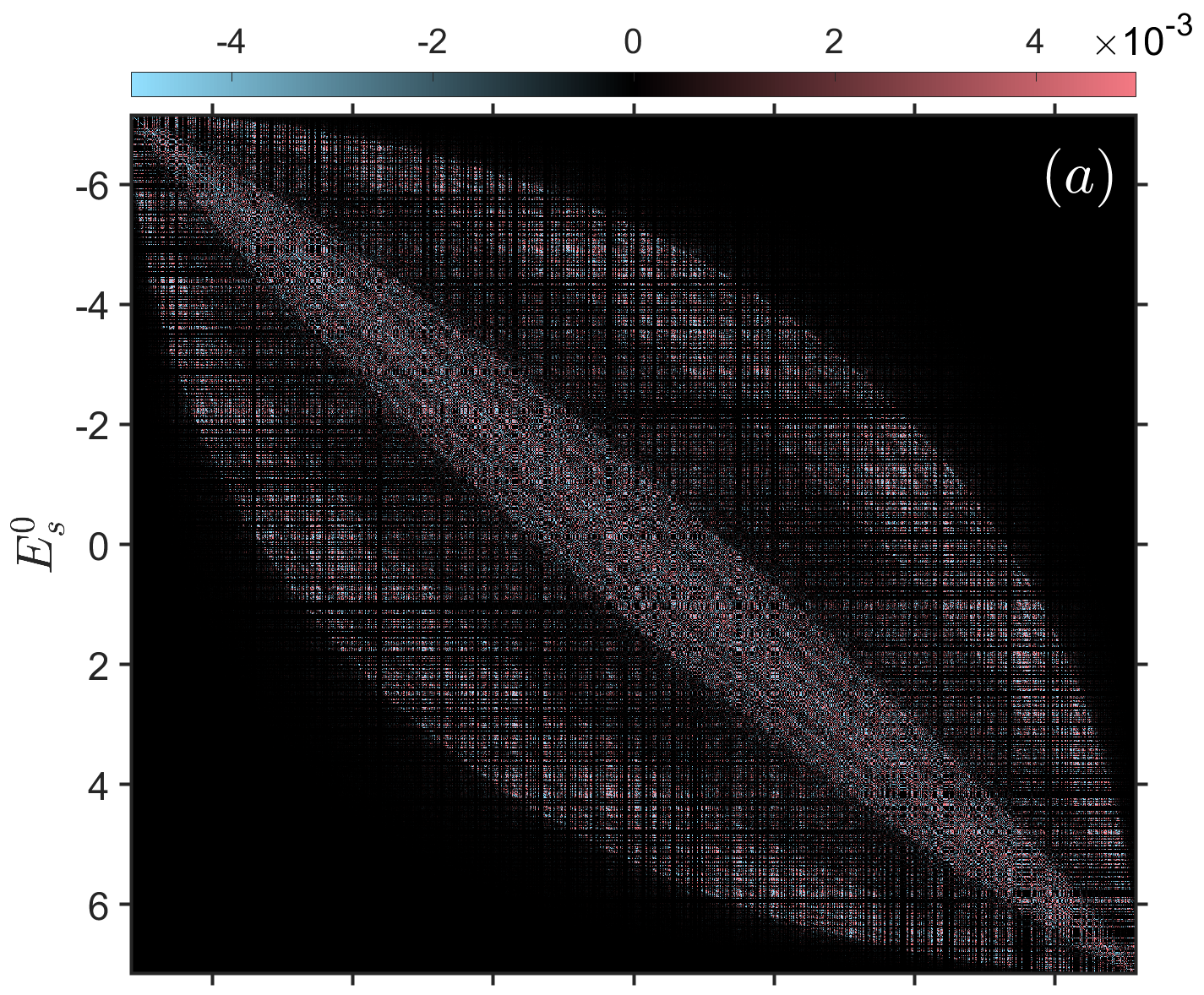}
    \includegraphics[width=0.962\linewidth]{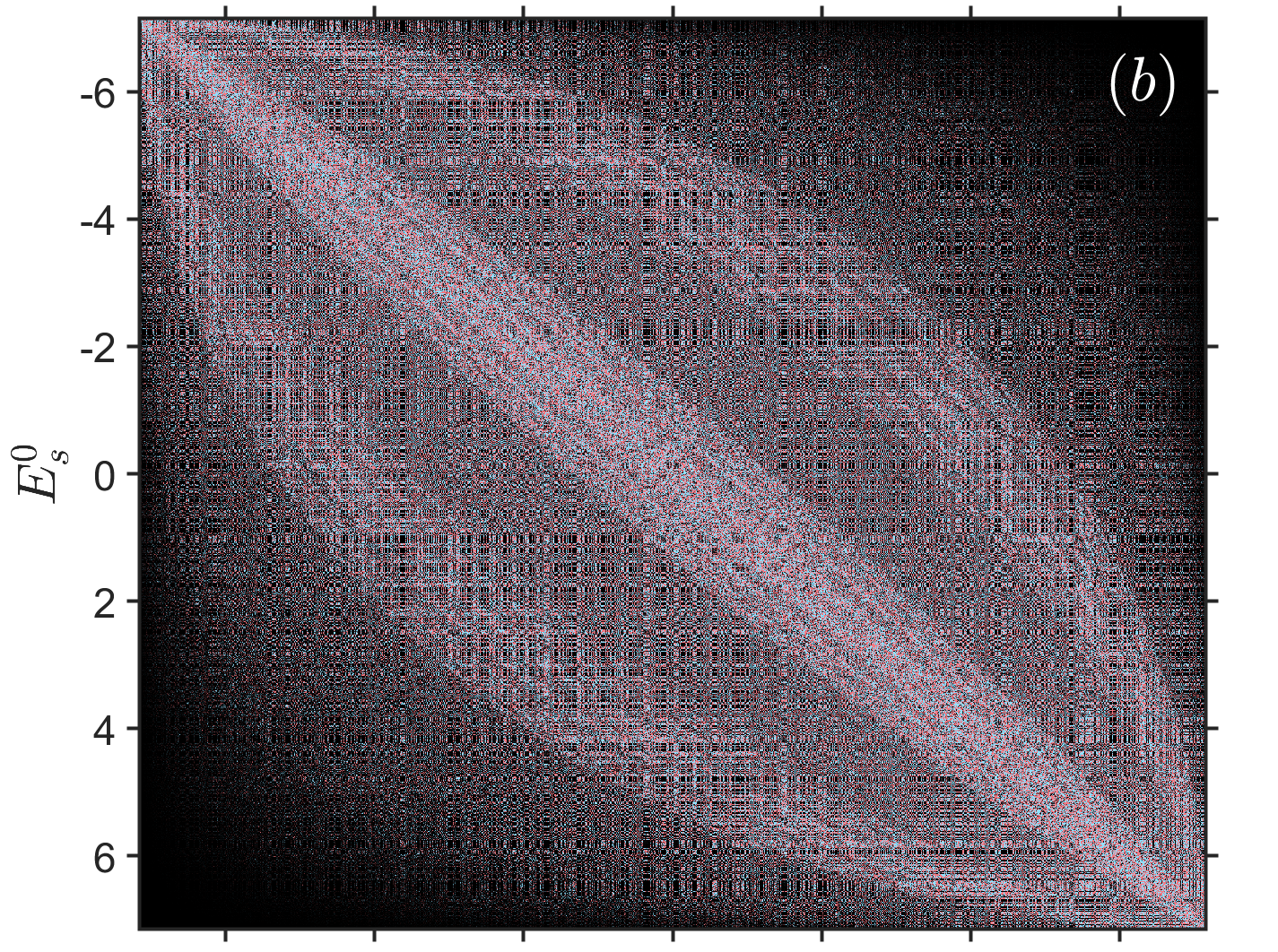}
    \includegraphics[width=0.962\linewidth]{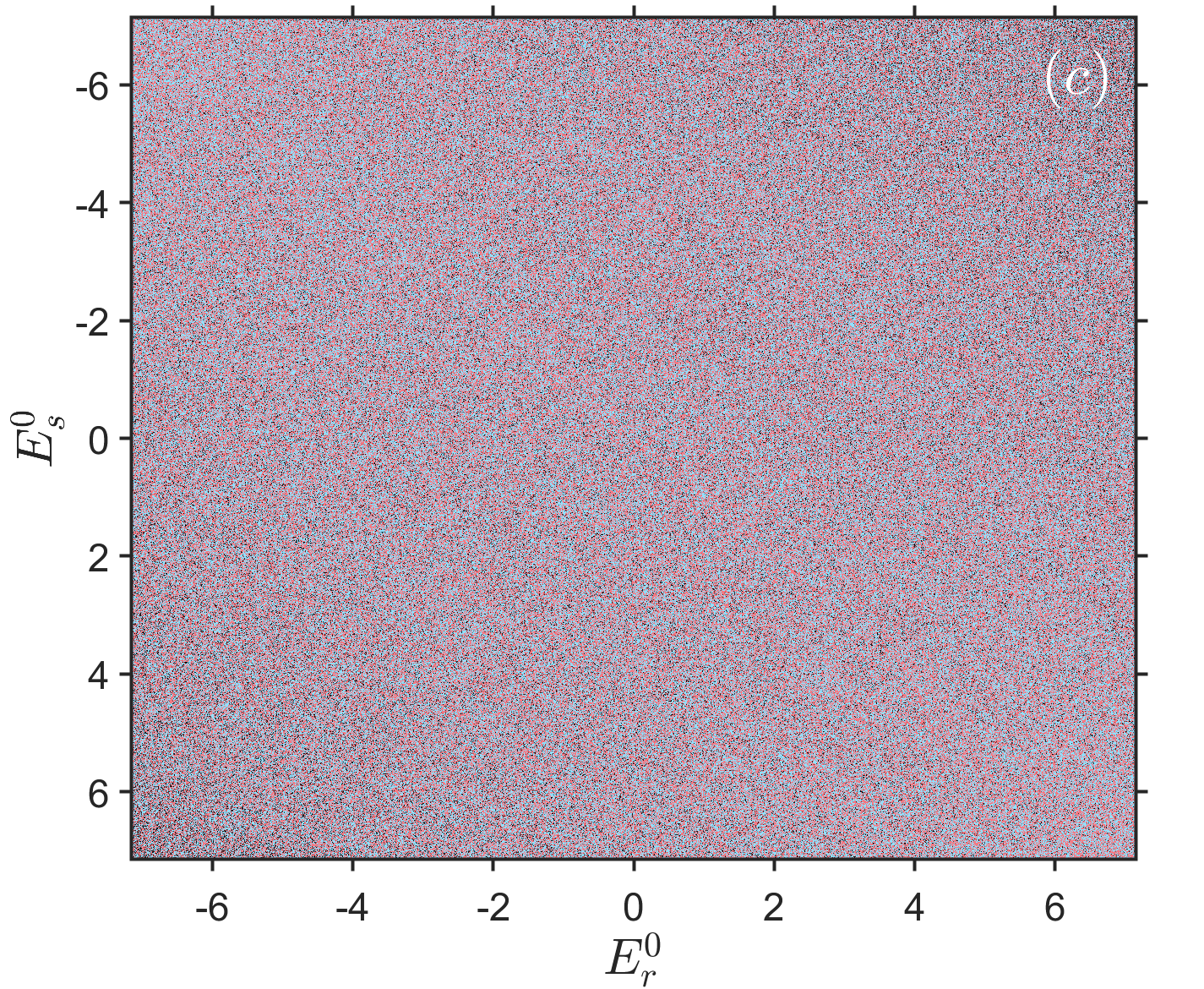}
    \caption{The matrix elements in the system-environment uncoupled basis $\ket{E_r^0}$.
    Specifically, panel (a) depicts the matrix elements $\bra{E_s^0}H_{\rm DIS}\ket{E_r^0}$,
    panel (b) illustrates the matrix elements $\bra{E_s^0}H^{\rm DIS}_{n=14}\ket{E_r^0}$,
    and panel (c) shows the matrix elements $\bra{E_s^0}H^{(R)}_{N_R=2^{13}\times 14}\ket{E_r^0}$.
    Here, the total spin number $N=14$.}
    \label{DIS_HShape}
\end{figure}

 For the purpose of understanding numerical simulations presented in
 previous sections, one meaningful question is as follows:
 Is there a special representation, which is of special relevance to the observable $O$,
 while, in which the two types of Hamiltonian discussed previously,
 show qualitatively different types of matrix structure?
 In this section, we show that a system-environment uncoupled basis is useful for this purpose.

 Let us consider a local observable $O$,
 which is for a central system $\mathcal{S}$.
 The rest of the total system is referred as the  environment $\mathcal{E}$.
 The total Hamiltonian is written as
\begin{equation}\label{Coupled_H}
    H=H_\mathcal{S}+H_\mathcal{E}+H_\mathcal{I},
\end{equation}
where $H_\mathcal{S}$ and $H_\mathcal{E}$ denote the Hamiltonians of $\mathcal{S}$ and
$\mathcal{E}$, respectively, determined under the weak-coupling limit, while $H_\mathcal{I}$
represents the interaction Hamiltonian between $\mathcal{S}$ and $\mathcal{E}$.
 The uncoupled system-environment Hamiltonian is written as
\begin{equation}\label{UnCoupled_H0}
    H_0=H_\mathcal{S}+H_\mathcal{E},
\end{equation}
with eigenvalues and eigenstates indicated as $E_r^0$ and $\ket{E_r^0}$, respectively,
in the increasing-energy order,
\begin{equation}
    H_0\ket{E_r^0}=E_r^0\ket{E_r^0}.
\end{equation}
 The set $\{\ket{E_r^0}\}$ constitutes the system-environment uncoupled basis.

In the DIS model, with $O=\sigma_x^7$,  the seventh
spin is taken as the central system $\mathcal{S}$ and the remaining spins as the
environment $\mathcal{E}$.
 Thus,
\begin{subequations}
    \begin{align}
        H_\mathcal{S}=&\frac{B_x}{2}\sigma^7_x,\\
        H_\mathcal{E}=&\frac{B_x}{2}\sum_{l\neq 7} \sigma^l_x + \frac{d_1}{2} \sigma^1_z + \frac{d_5}{2} \sigma^5_z \notag\\
        &+\frac{J_z}{2}\left(\sum_{l\neq 6,7}\sigma^l_z \sigma^{l+1}_z+\sigma^N_z \sigma^1_z\right).
    \end{align}
\end{subequations}

Fig.\ref{DIS_HShape} shows schematic plots for structures of
the matrix elements of $H_{\rm DIS}$ (Fig.\ref{DIS_HShape}(a)),
$H^{\rm DIS}_{n=14}$ (Fig.\ref{DIS_HShape}(b)), and $H^{(R)}_{N_R=2^{13}\times 14}$
(Fig.\ref{DIS_HShape}(c)) in the system-environment uncoupled basis $\ket{E_r^0}$.
It can been seen that significant elements of the realistic Hamiltonians
of $H_{\rm DIS}$ and $H^{\rm DIS}_{n=14}$
are confined to a few band-shaped areas, while those the $H^{(R)}_{N_R=2^{13}\times 14}$
spans almost over all of the basis states.
\color{black}

Note that since $\ket{E_r^0}$ are eigenstates of $H_0$, all off-diagonal elements of
$\bra{E_s^0}H\ket{E_r^0}$ come from the system-environment interaction $H_\mathcal{I}$.
Therefore, the extent of the region occupied by the Hamiltonian in the uncoupled basis
$\ket{E_r^0}$ actually reflects a characteristic of the interaction $H_\mathcal{I}$.

The above results show that, in the uncoupled basis $\ket{E_r^0}$,
the system-environment interaction $H_\mathcal{I}$ of a realistic quantum chaotic system
occupies merely a few band-shaped regions of the matrix.
This implies strong correlations within energy eigenstates and
gives an explanation to the exponential decay of the $f^2(E_i,E_j)$ function at
large $\Delta E$.
 Meanwhile, the envelop function is flat for $H^{(R)}_{N_R=2^{13}\times 14}$.

Finally, it is worth mentioning that perturbation theories offer a natural avenue for
connecting energy eigenstates with the system-environment uncoupled basis when treating
$H_\mathcal{I}$ as a perturbation.
For example, a perturbation theory, which gives convergent perturbation expansions
for part of eigenfunction even at strong perturbations \cite{WIC98,Jiaozi_JPA2018},
may be useful for future investigations concerning the relationship
between the matrix structure of $H_\mathcal{I}$ in the uncoupled basis and the behavior
of the $f(E_i,E_j)$ function.

\section{Discussions and Conclusions}\label{Sec_Conclusion}
In this paper,  we study the
distinctions between realistic quantum chaotic systems and systems described by RMT manifested in statistical properties of observables.
In particular, we investigate the structure of the envelope function \( f(E_i, E_j) \), defined within the framework of the ETH. Through numerical simulations, we observe the presence of exponential decay at large \( \Delta E \) for realistic systems, and its absence in systems described by RMT.
To further unveil the connection between the non-trivial structure of \( f(E_i, E_j) \) and the 'realisticity' of the system, we investigate two types of Hamiltonians where the degree of 'realisticity' can be tuned. Numerical results show that the non-trivial structure of \( f(E_i, E_j) \) becomes less prominent as the system becomes less realistic, eventually flattening out.

We provide a framework for understanding the underlying physics behind the numerical observations above, which is presented from the following three perspectives.
First, there are strong correlations within the energy eigenfunctions of realistic models. Second, realistic systems contain far fewer dynamical group elements in their Hamiltonians compared to random models. Finally, the special structures of realistic Hamiltonians are clearly reflected in the structures of interactions in the system-environment uncoupled representation.

Our results highlight the importance of correlations to the nontrivial shape of
the envelope function $f(E_i,E_j)$, which deviates in realistic quantum chaotic systems from those described by RMT. Quantitative study of the relationship between 
randomness of the system and structure of $f(E_i,E_j)$ will be a valuable direction for
future research. Additionally, it would be interesting to consider higher-order envelope
functions introduced in the context of the generalized ETH \cite{FoiniPRE,PhysRevLett.129.170603}.

\acknowledgments
This work was partially supported by the Natural Science Foundation of China under Grant
Nos.~12175222, 11535011, and 11775210.
J.W. acknowledges support from Deutsche Forschungsgemeinschaft (DFG), under Grant No.
531128043, and under Grant
No. 397107022, No. 397067869, and No. 397082825,
within the DFG Research Unit FOR 2692, under Grant
No. 355031190.

\bibliography{ETH-gFunc.bbl}

\end{document}